\documentclass{article} % For LaTeX2e
\usepackage{iclr2021_conference,times}

\usepackage{amsmath,amsfonts,bm}

\def\eqref#1{equation~\ref{#1}}
\def\1{\bm{1}}

\DeclareMathAlphabet{\mathsfit}{\encodingdefault}{\sfdefault}{m}{sl}
\SetMathAlphabet{\mathsfit}{bold}{\encodingdefault}{\sfdefault}{bx}{n}

\usepackage{hyperref}
\usepackage{url}
\usepackage{graphicx}
\usepackage{float}
\usepackage{amssymb}

\title{Network Compression for Machine-learnt Fluid Simulations}

\author{Peetak P. Mitra \\
University of Massachusetts, Amherst, USA \\
\texttt{\{pmitra\}@umass.edu} \\
\And
Dhananjay Kumar \& Nomit Jangid \& Ashwati Nambiar \\ 
The MathWorks, Inc., India\\
\And

Vaidehi Venkatesan \& Niccolo Dal Santo \& Shounak Mitra \\
The MathWorks, Inc., USA

\AND
Majid Haghshenas \& David P. Schmidt \\
University of Massachusetts, Amherst, USA \\
}

\iclrfinalcopy 
\begin{document}

\maketitle

\begin{abstract}

Multi-scale, multi-fidelity numerical simulations form the pillar of scientific applications related to numerically modeling fluids. However, simulating the fluid behavior characterized by the non-linear Navier Stokes equations are often times computational expensive. Physics informed machine learning methods is a viable alternative and as such has seen great interest in the community [refer to \cite{kutz2017deep, brunton2020machine, duraisamy2019turbulence} for a detailed review on this topic]. For full physics emulators, the cost of network inference is often trivial. However, in the current paradigm of data-driven fluid mechanics models are built as surrogates for complex sub-processes. These models are then used in conjunction to the Navier Stokes solvers, which makes ML model inference an important factor in the terms of algorithmic latency. With the ever growing size of networks, and often times overparameterization, exploring effective network compression techniques becomes not only relevant but critical for engineering systems design. In this study, we explore the applicability of pruning and quantization (FP32 to int8) methods for one such application relevant to modeling fluid turbulence. Post-compression, we demonstrate the improvement in the accuracy of network predictions and build intuition in the process by comparing the compressed to the original network state.
\end{abstract}

\vspace{-5mm}

\section{Introduction}

Overparameterized neural networks have recorded state-of-the-art performances in applications such as computer vision and natural language processing [\cite{neill2020overview, huang2018gpipe, brown2020language}] in recent years. These networks however, lead to a larger footprint for inference as well as huge memory and storage requirements. While there exists techniques to improve model inference by network compression, network pruning and quantization are emerging as important techniques that are algorithm and hardware-efficient [\cite{tanaka2019deep, arora2018stronger}]. Such network compression techniques can substantially reduce the computational demands of the inference when conducted in a fashion amenable to the hardware or hardware designed to explore sparsity [\cite{rocki2020fast}]. While there have been studies that explore pruning at network initialization [\cite{blalock2020state, frankle2020pruning}] to reduce cost of training itself, the scope of this work is limited to exploring feasibility of pruning methods for scientific applications, post-training.

While magnitude pruning [\cite{mozer1988skeletonization, hassibi1993second, janowsky1989pruning, reed1993pruning, han2015learning}] has shown impressive results, they involve significantly more computational cost to iteratively train and prune network parameters [\cite{tanaka2019deep}]. In addition, these methods in some cases undergo the phenomena of \textit{layer collapse}, rendering the network untrainable. Therefore a more appropriate approach in pruning is based on conserving gradient based scores at each neuron and layer of the network [\cite{tanaka2019deep, lecun1989optimal}]. In this study, the Synaptic Flow (SynFlow) pruning algorithm is used. Furthermore, the effect of quantization by storing information in fewer bits (FP32 to int8 precision) are explored. The compressed network performance is measured against a hold-out test dataset, taken from the same distribution as the training set. For scientific surrogate model applications, learning OOD performance is not critical. This is especially relevant to building models for complex physical processes that provide source terms to the larger numerical model as in the case of weather modeling [\cite{watt2021correcting}] and combustion [\cite{bode2021using}].

\subsection{Machine Learning for Fluids}

Scale bridging is a critical need in computational sciences, where the modeling community has developed accurate physics models from first principles, of processes at lower length and time scales that influence the behavior at the high scales of interest. However, it is not computationally feasible to incorporate all of the lower length scale physics directly into up-scaled models. This is an area where machine learning has shown promise, in building emulators of the lower length scale models which incur a mere fraction of the computational cost of the original higher fidelity models. Some related efforts can be found in \cite{portwood2019turbulence, portwood2020analysis, raissi2020hidden, ling2016reynolds, maulik2019subgrid,shankar2020rapid}. We extend this effort to emulate a data-driven turbulence sub-grid closure term for compressible flows relevant to modeling advanced propulsion systems - refer to Appendix C for more details. 

Most neural networks have inductive biases and \cite{mitchell1980need} argued that inductive biases constitute the heart of generalization and the key basis for learning itself [\cite{bworld}]. A key challenge of machine learning, therefore, is to design systems whose inductive biases align with the structure of the problem at hand. The effect of such efforts is not merely to endow the model with the capacity to learn key patterns, but also – somewhat paradoxically – to deliberately hamper the capacity of the model to learn other (presumably less useful) patterns, or at least to drive the model away from learning them. In other words, inductive biases stipulate the properties that we believe our model should have in order to generalize to future data; they thus encode our key assumptions about the problem itself. Unlike applications in computer vision where models are deployed on edge devices (ex. FPGAs), our choice of network is not limited to reducing number of parameters due to memory limitations. It is however dependent on reducing number of total operations - more details in Figure \ref{fig:Workflow}. Therefore our chosen network architecture is an eight-hidden layer feed-forward dense neural network that has similar inductive biases of our problem of interest -- i.e. to determine functional relationship between parameters to predict a series of outputs -- and lower number of operations compared to a spatial relationship aware Convolutional Neural Network (CNN).

In this study, we explore a previously well trained network developed expressly for this task. In order to automate the process of network design, a Bayesian optimization based autoML method is used - more details can be found in previous work [\cite{mitra2020effectiveness}]. Once the best performing setting for network architecture and hyper-parameters are identified, they are coupled to an open source PDE solver, OpenFOAM [\cite{jasak2007openfoam}]. While Appendix A and C describe the approach briefly, for a detailed review refer to the original manuscript [c.f. \cite{mitra2021analysis}].

\subsection{The need for network compression}

The machine-learnt subgrid model, now coupled to the C++ solver, OpenFOAM [\cite{jasak2007openfoam}], is used for inference - in a workflow schematic as shown in Appendix A. This inference is obtained at each node point, for each time-step. Typically for a engineering level fluid simulation, the number of meshpoints and timesteps are in the order of a few million. This essentially means the number of network inference function calls are of the same order. For a deep neural network that amounts to $O*T*N$ operation calls, where $O$ is the number of matrix operations, $T$ is the number of overall simulation time-steps and $N$ are the total number of node points. Even if the well-trained surrogate model alleviates the need to numerically solve for PDEs for specific tasks, these repetitive costs accumulate and become expensive for reasonably sized problem. This also leads to a latency between different components of the numerical solver that incorporates the inference calls, which slows down overall computation.

A useful strategy in alleviating this would be to use a reduced precision, that will improve the arithmetic and memory bandwidths for inference. For example, half-precision math throughput in recent GPUs is 2× to 8× higher than for single-precision [\cite{al2020tensor}]. In addition to speed improvements, reduced precision formats also reduce the amount of memory required for training and inference. Therefore, pruning and quantization can play an important role in network compression and improving the inference throughput.

\section{Network Compression}
\subsection{Pruning Approach}
\label{gen_inst}

Conventional pruning algorithms assign scores to parameters in neural
networks after training and remove the parameters with the lowest scores. Popular scoring
metrics include weight magnitudes, its generalization to multi-layers, first-
and second-order Taylor coefficients of the training loss with respect to the parameters, and
more sophisticated variants. While these pruning algorithms can indeed compress neural
networks at test time, there is no reduction in the cost of training.  On the other hand the SynFlow approach [\cite{tanaka2019deep}] uses synaptic saliency as a metric for pruning defined as:

\begin{equation}
    S(\theta) = \frac{\partial \mathrm{R}}{\partial \theta} \odot \theta ;  S^{in} = \frac{\partial \mathrm{R}}{\partial \theta^{in}} \odot \theta^{in} ;    S^{out} = \frac{\partial \mathrm{R}}{\partial \theta^{out}} \odot \theta^{out}
\end{equation}

where $\mathrm{R}$ is a scalar loss function of the output $y$ of a feed-forward network parameterized by $\theta$. When
$\mathrm{R}$ is the training loss $\mathrm{L}$, the resulting synaptic saliency metric is equivalent (modulo sign) to  - $\frac{\partial \mathrm{L}}{\partial \theta} \odot \theta$, the score metric used in skeletonization \cite{lecun1989optimal}, one of the first network pruning algorithms. The resulting score metric is also closely related to $|\frac{\partial \mathrm{R}}{\partial \theta} \odot \theta|$. In the SynFlow approach, \cite{tanaka2019deep}, the neuron-wise conservation of synaptic saliency is maintained such $S^{in} = S^{out}$.

\subsection{Quantization}

Post-training quantization is an important tool for optimizing deep learning models, as it helps accelerate inference. To accomplish this goal of int8 quantization, the Deep Learning Quantizer app within \cite{MATLAB:2020a} is used. The important aspects of quantization is the precision loss due underflow and the overflow of the stored information. Appendix B discusses this briefly.

\section{Results}
\label{headings}

We use two thresholds to compare the performance of the pruning algorithm. First is a threshold of pruning 50\% parameters and the second a threshold of pruning 90\% parameters. These thresholds although somewhat arbitrary, will demonstrate the possible compression of the network and the effect on the inference performance. The pruning is done on all the network layers, $fc_1$ to $fc_9$ in an iterative fashion. The resulting outcome from each iteration is described in detail in Appendix B while the final state is discussed in Figure 1. Tables \ref{tab:PrunedParameters} and \ref{tab:PruningMSE} shows the relative number of parameters pruned in each setting and the corresponding Mean Squared Error (MSE) measure on the unseen validation dataset, randomly sampled from the original dataset. Additionally, the effects of quantization were explored and reported in Table \ref{tab:PruningMSE}. It follows that for both the pruning thresholds, the network performance (MSE) improves over the baseline, determined by the original network MSE on the same data. This outcome is not entirely surprising, as overparameterization and the propensity of deep neural networks to learn noise for regularization, and better generalization can lead to lower baseline scores [\cite{bao2019using}]. Pruning can often remove the learnt 'noise' which improves performance at the cost of robustness and generalizability. Figure \ref{fig:Histogram} shows the histogram of the original network and the min/max range of the data is within the overflow and underflow range of the int8 precision (-128 to 127) which leads to insignificant precision losses. 

For scientific surrogate models, while the network inference is key in accelerating compute, the most important metric is the accuracy of the model output. This gains relevance as the ML models are coupled to non-linear PDE solvers which makes the overall system extremely sensitive to the accuracy of these data-driven sub-models. The 50\% pruned network shows remarkable consistency with the original network characteristics (see Figure \ref{fig:SVDSSIM}) with the added advantage of an atleast 5x speedup [refer Table \ref{tab:netInfer}]. The loss in the saliency score is about an order of magnitude for the 50\% threshold, compared to three orders of magnitude for the 90\% pruning (refer Table \ref{tab:ScoreSum}). This makes the former a stronger candidate for pruning thresholds for the given problem. 

\begin{figure}
    \centering
    \includegraphics[width=0.24\linewidth]{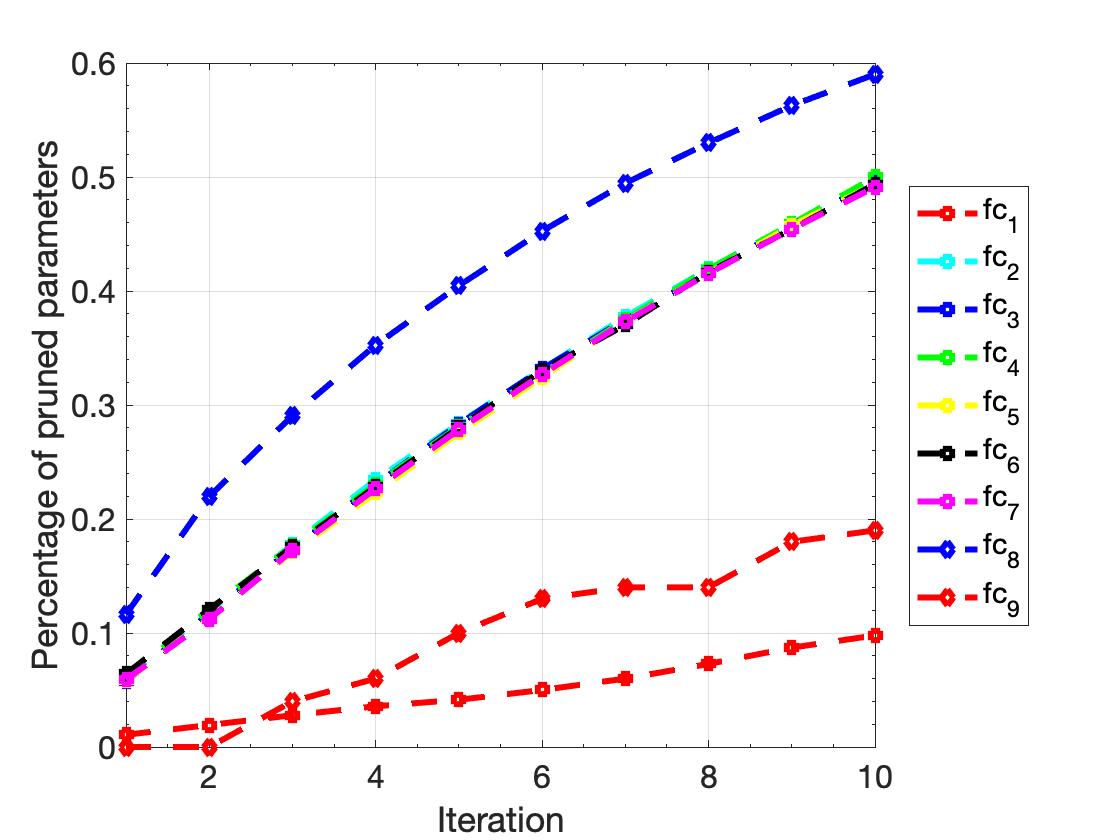}
    \includegraphics[width=0.24\linewidth]{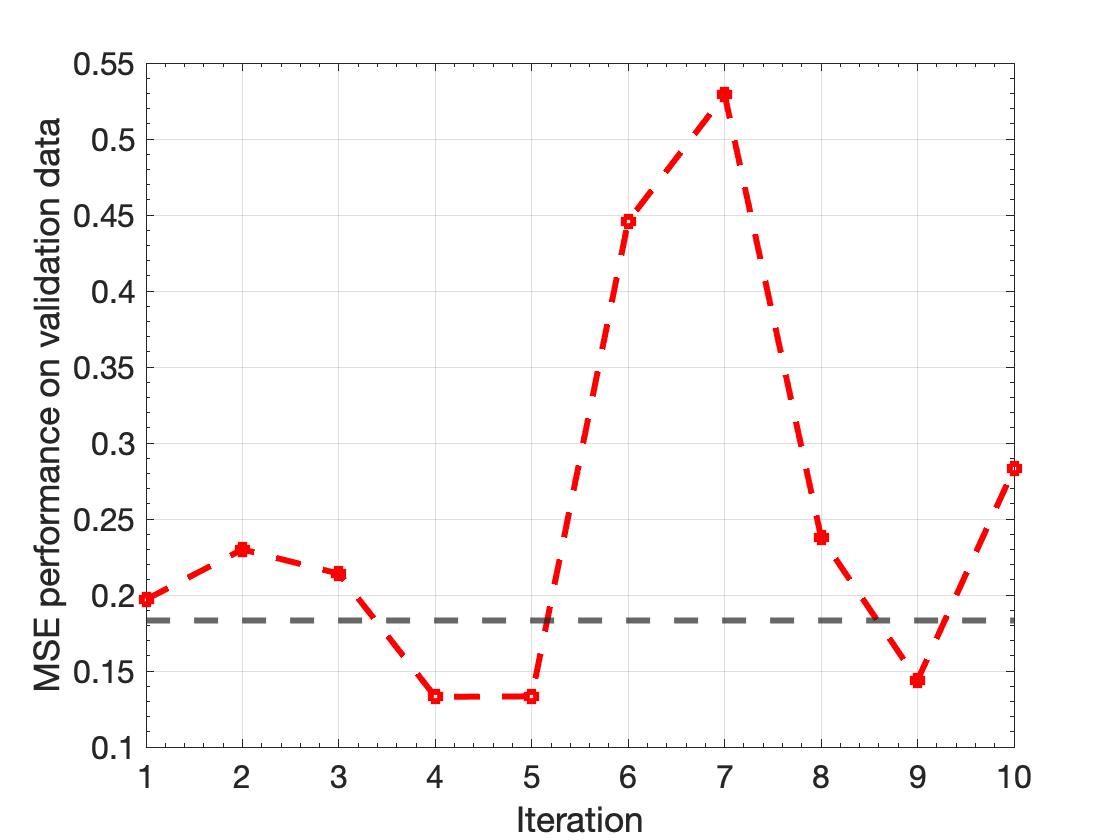}
    \includegraphics[width=0.24\linewidth]{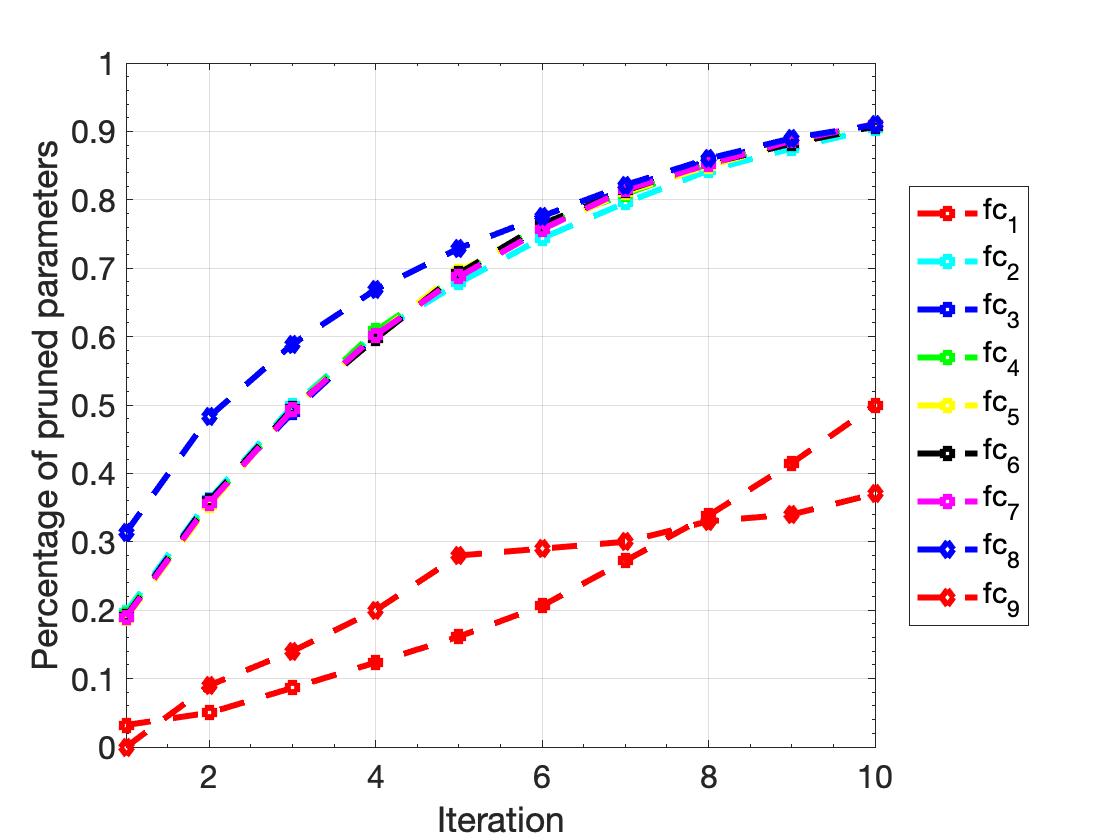}
    \includegraphics[width=0.24\linewidth]{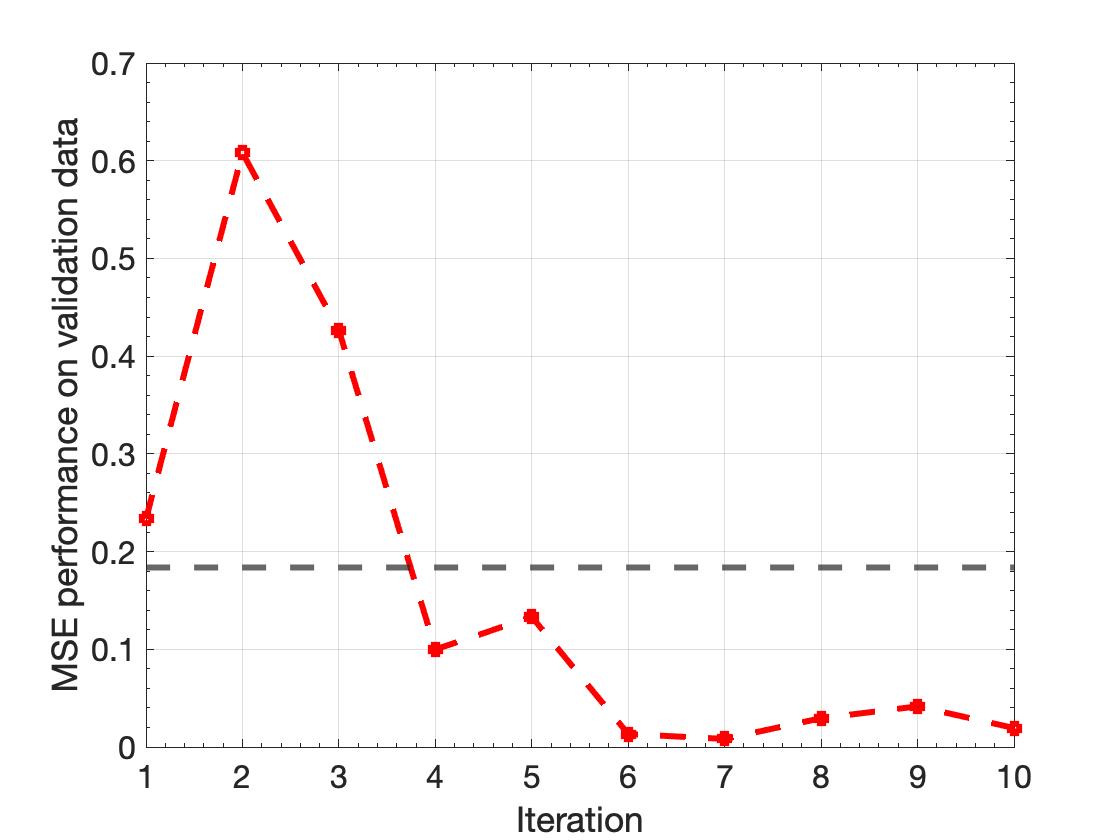}
    \caption{[L-R] As the pruning iteration increases, the larger share of parameters are pruned from deeper layers preventing layer collapse. Pruning 50\% parameters from the network shows a slight improvement over original MSE (marked in dotted horizontal line) of the full network. Pruning 90\% parameters show significant improvement over original MSE baseline. This can be attributed to the removal of the noise in the learning process. The gain in speedup is compensated by the loss in generalizability for these models. However, for many surrogate modeling applications this may be acceptable.}
    \label{fig:50_90pc}
\end{figure}

\subsection{Exploring similarities within low rank representation}

\begin{figure}
    \centering
    \includegraphics[width=0.3\linewidth]{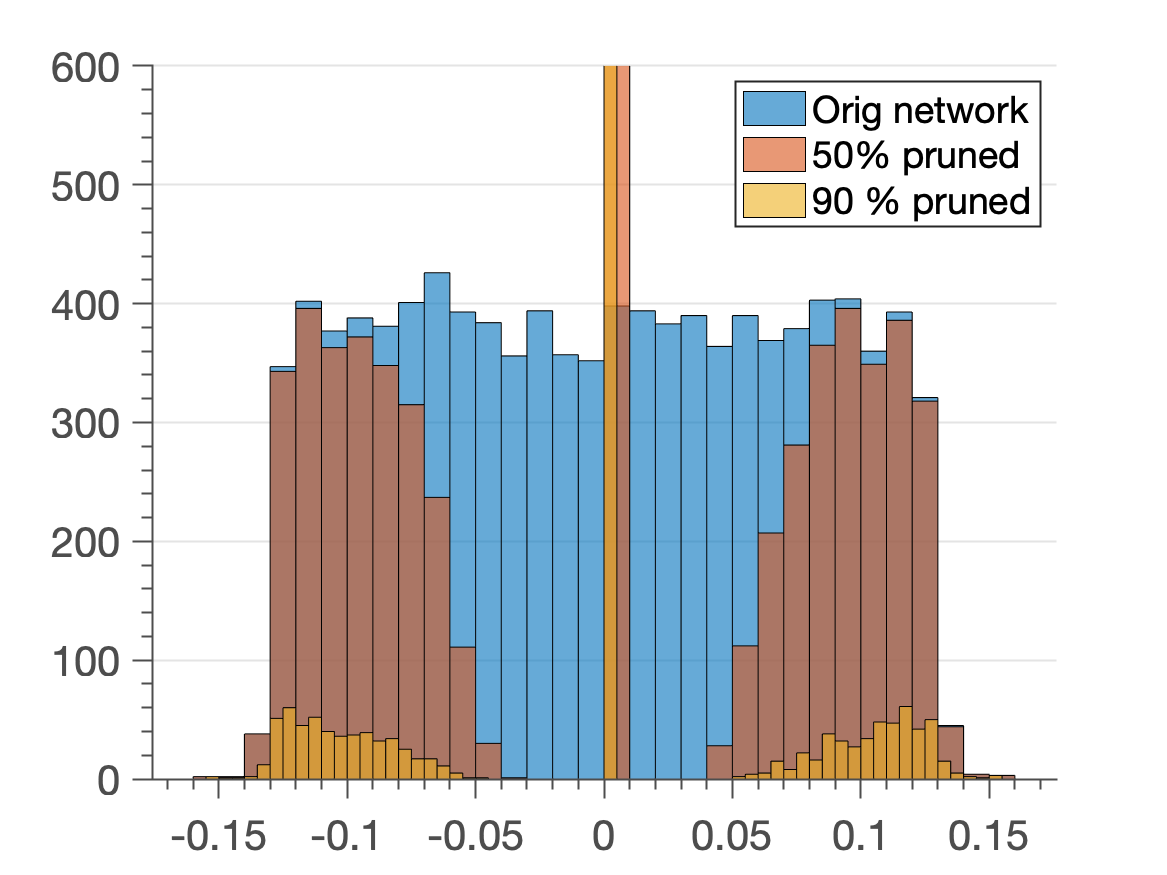}
    \includegraphics[width=0.3 \linewidth]{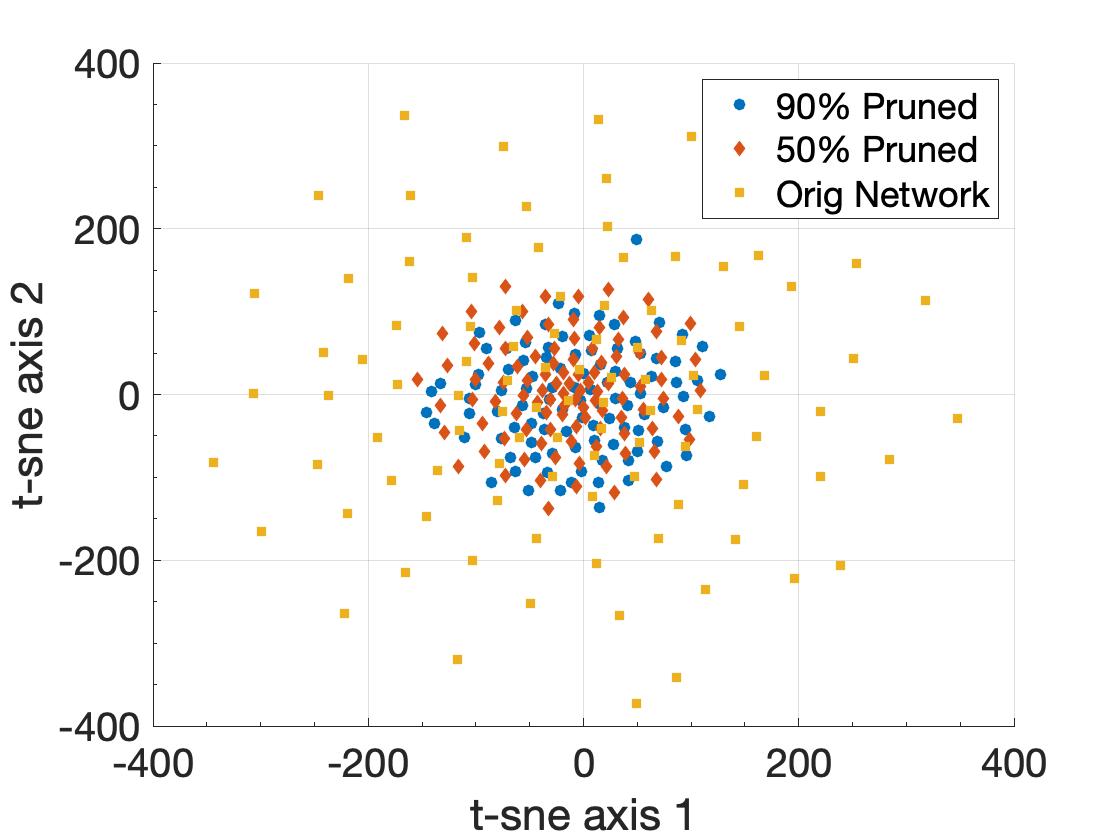}
   
    \caption{Comparison of the layer connections for $fc_4$ between the original and pruned networks shows the pruning is done primarily for very low values of the weights, preserving the representational value of the network and removing 'noise'. The low-dimensional representation, obtained from t-SNE, shows the overlapping function-space similarities in the original to the compressed networks, which explains the generally good inference performance for latter with fewer parameters.}
    \label{fig:Histogram}
\end{figure}

To better understand the predictions, we plot the histogram of the original network, and the two pruned networks for the layer 4 connections ($fc_{4}$), in Figure \ref{fig:Histogram}, as it is a representative sample of the neural connections in the whole network. The result shows that the pruned networks only retain the most consequential information, i.e. at the edges of the histogram, whereas the noise in the form of small weight values are not retained. This supports the discussion in the previous section. Visualizing the distribution of network weights in a low-dimensional space for the same layer, using t-SNE [\cite{van2008visualizing}], shows a similar trend of overlapping regions of similarity, whereas the original network also retains some sparse noise. Overall, for the cost of pruning, and the efficiency gained the 50\% pruned network overall appears a good candidate for the network compression.

\section{Conclusions}

Network pruning, or the compression of neural networks by removing parameters, has been an
important subject both for reasons of practical deployment  and for theoretical
understanding of artificial [\cite{arora2018stronger}] and biological [\cite{liu2018progressive}] neural networks.
Pruning shows an immediate effect on the network predictions by improving the performance of the network on the original dataset in Figure \ref{fig:50_90pc}. This is not a surprising result as previous studies [\cite{tanaka2019deep, arora2018stronger}] have reported minor improvements in accuracy during post-training pruning exercises. These pruned models are expected to have comparatively poorer generalizability characteristics. However, for applications related to building surrogate models the networks are expected to have good performance for datasets from within the training distribution, therefore sacrificing generalizability to achieve faster network inference might be relevant to many applications of scientific interest relevant to climate/weather modeling and building robust digital twins, among others.

%\subsubsection*{Author Contributions}

%\subsubsection*{Acknowledgments}

\bibliography{iclr2021_conference}
\bibliographystyle{iclr2021_conference}

\appendix

\section{Overall approach}

Fluid turbulence is a multi-scale phenomenon and is an essential component of modeling engineering-relevant flows. While solving the full Navier Stokes using Direct Numerical Simulation (DNS) results in the most accurate representation of the complicated, non-linear, non-local, multi-scale phenomenon, DNS is often computationally intractable. Engineering level solutions based on Reynolds Averaged Navier-Stokes (RANS) and Large Eddy Simulations (LES) alleviate this issue by resolving the larger integral length scales and modelling the smaller unresolved scales by introducing a linear operator to the Navier-Stokes equation to reduce the simulation complexity. These models however suffer from the curse of turbulence closure.  The linear eddy-viscosity model represents one of the most popular methods for Reynolds stress closure for two-equation RANS as well as Smagronisky-LES models \cite{germano1991dynamic, smagorinsky1963general}. However, these approximates models are commonly phenomenological/heuristic in nature and thus require fitting to high fidelity DNS datasets for idealized flows in anycase \cite{portwood2019turbulence}.

\begin{figure}[H]
    \centering
    \includegraphics[width=0.4\linewidth]{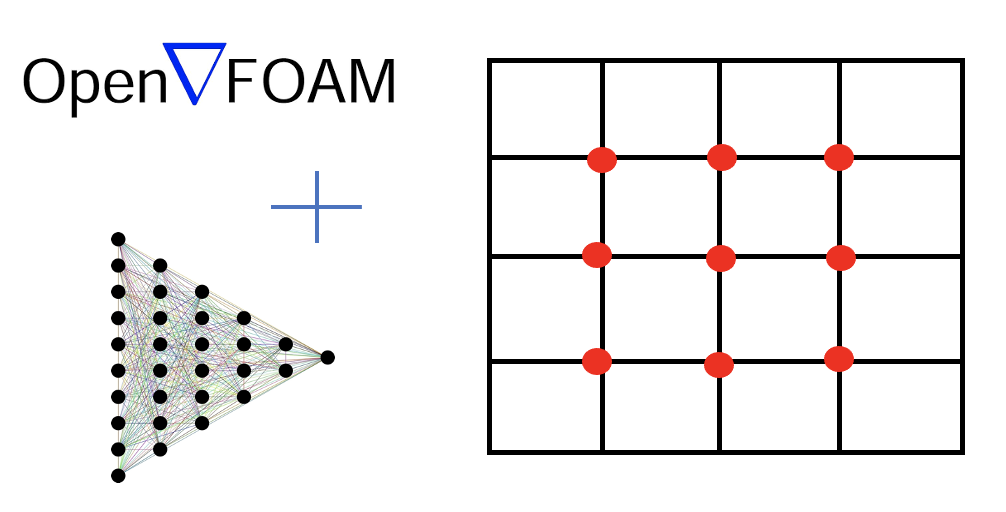}
    \caption{The current modeling paradigm includes using the neural network as a surrogate model within the non-linear PDE solver. The neural network is used as an inference engine within the OpenFOAM \cite{jasak2007openfoam} code and the inferences scales with the nodes (indicated here as the red dot on a simplified mesh structure) at every iteration of the simulation. A typical CFD problem involves millions of mesh points and thousands of timesteps before a solution is reached, thereby exacerbating the challenge in slow network inference.}
    \label{fig:Workflow}
\end{figure}

In the original work, a data-driven turbulence closure model is built. The full details can be found in the original work [\cite{mitra2021analysis}]. Once fully trained the network is then converted to C++ source code using MATLAB [\cite{MATLAB:2020a}] Code generator tools, which is then fully integrated into the OpenFOAM Navier-Stokes solver. During runtime for the coupled framework [see Figure \ref{fig:Workflow}], the network inference function along with various libraries in OpenFOAM are called at each timestep and at each node. Therefore the latency in the network inference function call, potentially slows down the overall simulation. This paradigm is similar to applications in weather forecasting as well as combustion [\cite{bode2021using}].

\section{Network compression, additional results}

The final state for the layer wise pruning, in Table \ref{tab:PrunedParameters}, indicates the shallower layers lose fewer parameters while the deeper layers lose more parameters as a percentage of the total per-layer connections. This is consistent with the SynFlow algorithm and essentially ensures there is no layer collapse in the network.

\begin{table}[H]
    \centering
    \begin{tabular}{|c|c|c|c|}

\hline
   \textbf{Layer} & \textbf{Total parameters}  & \textbf{50\% pruned network} & \textbf{90\% pruned network} \\
   \hline
    $fc_1$ & 1400 & 137 & 700  \\
    \hline
    $fc_2$ & 10000 & 4949 & 9045 \\
    \hline
    $fc_3$ & 10000 & 4956 & 9099 \\
    \hline
    $fc_4$ & 10000 & 5000 & 9082 \\
    \hline
    $fc_5$ & 10000 & 4938 & 9109 \\
    \hline
    $fc_6$ & 10000 & 4938 & 9074 \\
    \hline
    $fc_7$ & 10000 & 4912 & 9108 \\
    \hline
    $fc_8$ & 10000 & 5901 & 9096 \\
    \hline
    $fc_9$ & 100 & 19 & 37 \\
    \hline
    MSE & 0.1874 & 0.133 & 0.029 \\
    \hline
\end{tabular}
    \caption{The total number of pruned parameters for each threshold (50\% and 90\%) at the end of the final iteration. The MSE performance improves over the baseline (original network).}
    \label{tab:PrunedParameters}
\end{table}

\subsection{Computational footprint of the compressed networks}

Apart from the discussions regarding the compressed network inference performance in terms of validation dataset accuracy and the speed up, Table \ref{tab:PruningMSE} discusses the footprint of the original network compared to the compressed networks in terms of space occupied by the weights and the run time memory. As expected, with a higher compression ratio, the memory occupied by weights and the runtime memory requirements reduce.

\begin{table}[H]
    \centering
    \begin{tabular}{|c|c|c|c|c|c|}
    \hline
        \textbf{Network} & \textbf{\# Weights} & \textbf{Memory (Weights)} & \textbf{\# Operations} & \textbf{Runtime memory} & \textbf{MSE*}  \\
        \hline
        Original & 71500 & 0.27 MB & 142199 & 1.63 MB & 0.1840 \\
        \hline
        \textbf{Unpruned Quantized} & 71500 & 0.14 MB & 142199 & 0.55 MB & \textbf{0.1680} \\
        \hline
        50\% Pruned & 35750 & 0.14 MB & 70699 & 0.8 MB & 0.2694 \\
        \hline
        \textbf{50 \% Pruned Quantized} & 35750 & 0.07 MB & 70699 & 0.27 MB & \textbf{0.1053} \\
        \hline
        \textbf{90\% Pruned} & 14300 & 0.05 MB & 27799 & 0.32 MB & \textbf{0.0339} \\
        \hline
        \textbf{90 \% Pruned Quantized} & 35750 & 0.03 MB & 70699 & 0.11 MB & \textbf{0.0518} \\
        \hline
    \end{tabular}
    \caption{The effect in terms of runtime memory and space occupied by network weights is inversely proportional to the compression ratio. The MSE is tested on an unseen validation dataset containing 10,000 samples. The \textbf{bolded} networks indicate better than baseline performing settings.}
    \label{tab:PruningMSE}
\end{table}

\subsection{Conservation of saliency score}

\begin{figure}[H]
    \centering
    \includegraphics[width=0.32\linewidth]{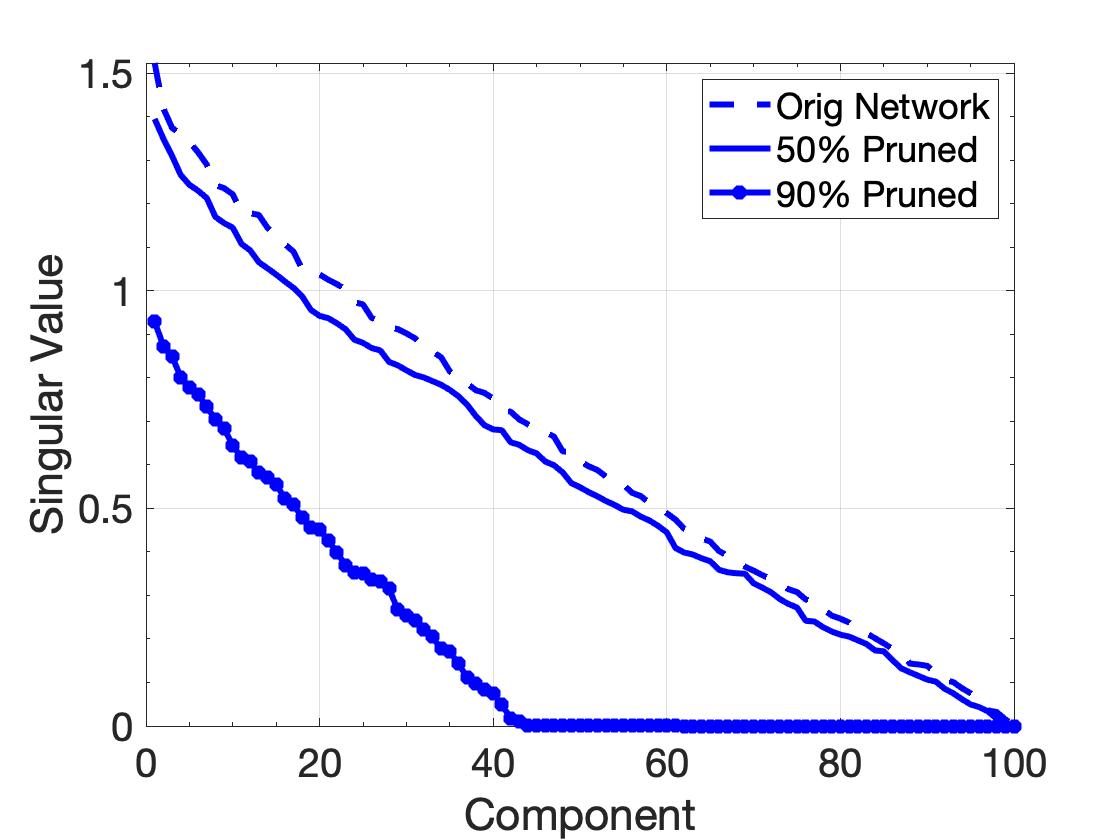}
    \includegraphics[width=0.32\linewidth]{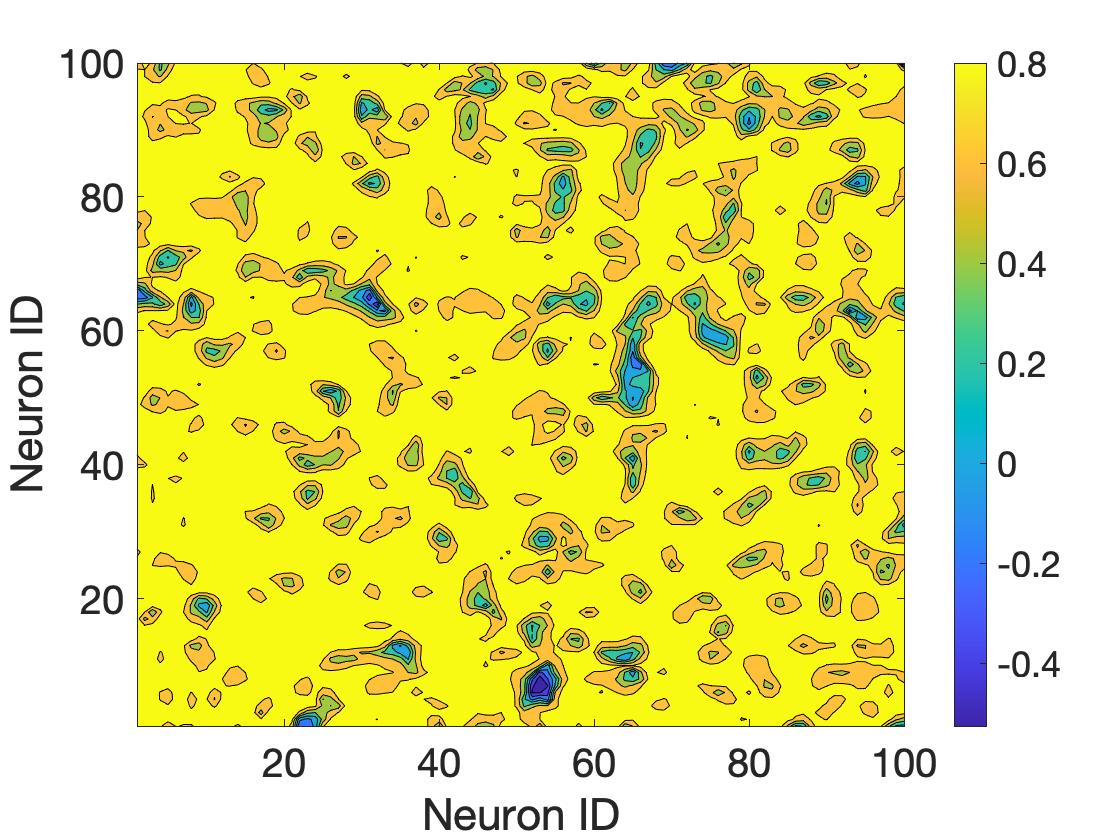}
    \includegraphics[width=0.32\linewidth]{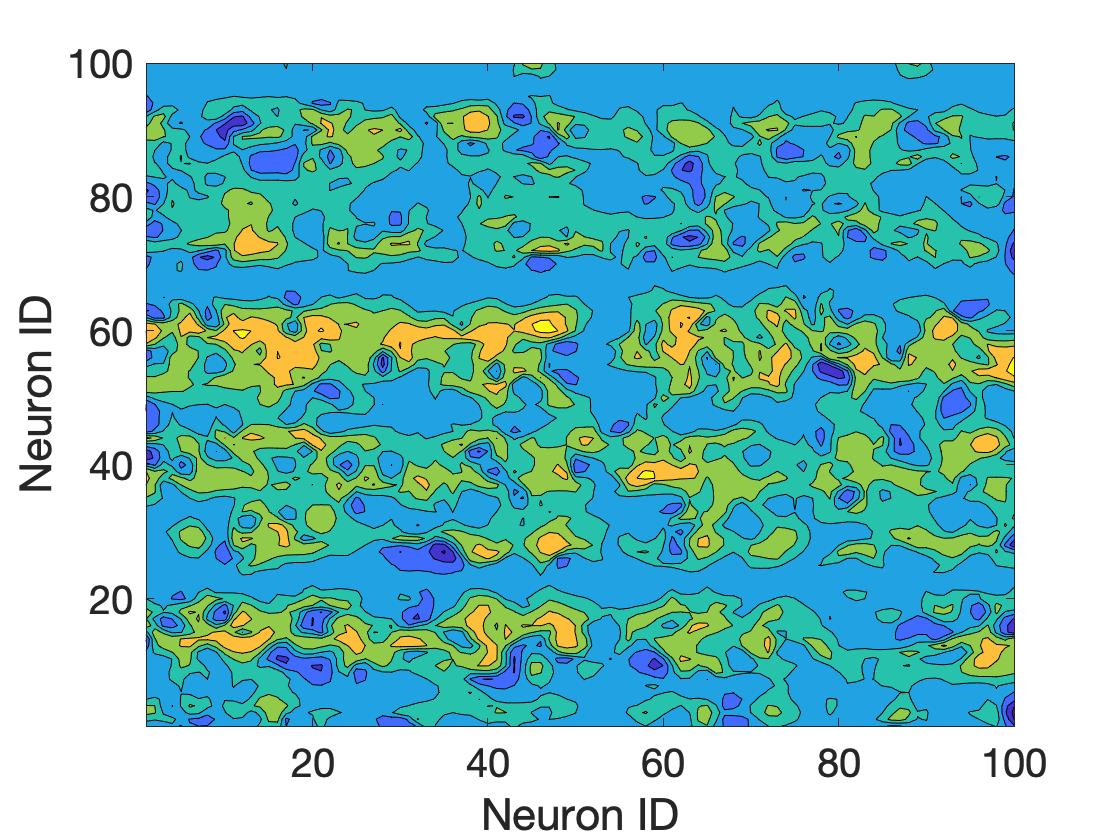}
    
    \caption{The similarity in the information retained as shown from the singular value plot on the left and the structural similarity between original-50\% pruned networks and original-90\% pruned networks reveal important characteristics. The 50\% pruning effectively retains a lot of the information in the origina network, in a compressed space.}
    \label{fig:SVDSSIM}
\end{figure}

The conservation of the saliency score (refer equation 1) is shown in Table \ref{tab:ScoreSum} for both pruning thresholds. The 50\% pruned network shows a loss of an order of magnitude compared to three for the higher compression ratio setting. This is consistent to the singular value observation seen in Figure \ref{fig:SVDSSIM}.

\begin{table}[H]
    \centering
    \begin{tabular}{|c|c|c|c|c|c|c|c|c|c|c|}
    \hline
    \textbf{Network/Epoch} & 1 & 2 & 3 & 4 & 5 & 6 & 7 & 8 & 9 & 10\\
    \hline
        \textbf{50\% pruned} & $1.09e^8$ & $1.06e^8$ & $9.92e^7$ & $8.80e^7$ & $7.52e^7$ & $6.20e^7$ & $4.98e^7$ & $3.90e^7$ & $3.00e^7$ & $2.27e^7$  \\
        \hline
        \textbf{90\% pruned} & $1.09e^8$ & $8.40e^7$ & $4.26e^7$ & $1.69e^7$ & $6.09e^6$ & $2.46e^6$ & $1.13e^6$ & $5.20e^5$ & $2.45e^5$ & $1.10e^5$  \\
        \hline
    \end{tabular}
    \caption{The saliency score for both pruned networks show that the 50\% pruning conserves much of the original information, consistent with observations in Figure \ref{fig:SVDSSIM}.}
    \label{tab:ScoreSum}
\end{table}

\subsection{Inference speedup}

Among the primary motivation to explore the feasibility of network compression is to improve network inference. Table \ref{tab:netInfer} shows the performance of the inference engine in a simulation-relevant environment. Since the quantization involves storing information in lower bits, from FP32 to int8 precision, an additional 3-4x speedup is expected for the networks.

\begin{table}
    \centering
    \begin{tabular}{|c|c|c|c|c|}
    \hline
     \textbf{Network} & \textbf{Sample size: 1000} & \textbf{Sample size: 1000000} & \textbf{Speedup}  & \textbf{Overall speedup} \\
    \hline
    Original & 0.19 s & 1.5 s & 1x & 3-4x\\
    \hline
    50 \% pruned & 0.14 s & 1.1 s & 1.5x & 4.5-6 x \\
     \hline
    90 \% pruned & 0.06 s & 0.47 s & 3x & 9-12x\\
    \hline
    \end{tabular}
    \caption{The inference speed-up post pruning for different data sizes shows up to 5-10x overall improvement in network inference without appreciable loss in generalizability. The results from different samples are shown here to demonstrate consistency in inference throughput.}
    \label{tab:netInfer}
\end{table}

\section{Physics of the problem}

The LES-filtered governing equations (using Favre-averaging) for the balance of mass and momentum are as below:
\begin{equation}
    \frac{\partial \bar{\rho}}{\partial t} + \frac{\partial }{\partial x_{j}} (\bar{\rho}\tilde{u_{j}}) = 0
\end{equation}
\begin{equation}
    \frac{\partial(\bar{\rho}\tilde{u_{i}})}{\partial t} + \frac{\partial \bar{\rho} \tilde{u_{i}}\tilde{u_{j}}}{\partial x_{j}} = \frac{\partial}{\partial x_{j}}[\bar{\rho}\bar{\nu}(\frac{\partial \tilde{u_{j}}}{\partial x_{i}}+\frac{\partial \tilde{u_{i}}}{\partial x_{j}}) -\frac{2}{3}\bar{\rho}\bar{\nu}\frac{\partial \tilde{u_{k}}}{\partial x_{k}}\delta_{ij}-\rho\tau_{ij}^{sgs}]-\frac{\partial \bar{p}}{\partial x_{i}} + \bar{p}g_{i} 
\end{equation}

where $u$ represents the velocity, $p$ is the pressure, $\rho$ the fluid density, $\nu$ the dynamic the viscosity and $\tau$ the subgrid stress term. The effect of the sub-grid scale appears on the right hand side of the governing equations through the sub-grid scale stresses, $\tau_{ij}$, which are modelled using the Boussinesq approximation \cite{spiegel1960boussinesq}, and the assumption by Smagorinsky that the smallest scales are isotropic \cite{smagorinsky1963general}. Based on Prandtl mixing length theory, the subgrid viscosity can be derived in terms of characteristic length and one velocity scale \cite{malalasekera2007introduction} as follows, therefore helping to close the Reynolds stress term
\begin{equation}
    {\tau_{ij}}^{sgs} - \frac{1}{3}{\tau_{kk}}^{sgs} \delta_{ij} = - \mu_{sgs}S_{ij}
\end{equation}

\begin{equation}
    \mu_{sgs} = \rho(C_{s}\Delta)^2|\bar{S}|
\end{equation}

where the superscript $sgs$ stands for subgrid scale terms, $\Delta$ is the filter width, and $C_s$ is the Smagorinsky constant, and $S_{ij}$ is the strain rate tensor and is calculated by taking off-diagnal gradients of velocities. In conclusion, the above approximation for the eddy viscosity assumes that changes in the resolved fields are slow, so that subgrid eddies can adjust themselves quickly to the rate-of-strain tensor. Thus, a closure based on a single constant is not universally true and the constant value may have to be adjusted \cite{meldi2011smagorinsky}, based on fitting the model parameters to high-fidelity data. Since some form of data-fitting is needed to optimize the subgrid scale model parameters even for this simplified approach, one can envisage a purely data-driven method to optimally approximate this changing constant based on large scale resolved terms, motivating our approach. More details about the LES implementation as well as the accuracy of results is reported in \cite{ribeiro2020large}.

In this work we aim to derive a functional relationship between the large scale resolved flow features and the sub-grid scale unresolved terms, and specifically to approximate the subgrid scale viscosity

\begin{equation}
    \mu_{sgs} = f(Re_c, S, \Omega, \triangledown, K, Y) 
\end{equation}

where $Re_{c}$ is the Cell Reynolds number, $S$ is the Strain-rate tensor and has six components, $\Omega$ is the rotation-rate tensor, and has three components, $\triangledown K$ is the Kinetic energy gradient, and $Y$ is a non-dimensional term that is a measure of the mesh resolution. The non-dimensionalized input features are chosen in order to impose Galliean-invariance \cite{speziale1985galilean, ling2016reynolds}. This functional mapping between a set of inputs to an output corresponds to the inductive biases of the dense fully connected network, hence the choice of the said architecture.

\section{Additional analysis}

\subsection{Structural Similarity}

Structural Similarity Index Measure (SSIM) is a perception-based model that considers image degradation as perceived change in structural information, while also incorporating important perceptual phenomena, including both luminance masking and contrast masking terms \cite{wang2004image}. Structural information is the idea that the pixels have strong inter-dependencies especially when they are spatially close. These dependencies carry important information about the structure of the data in the matrix. This metric applied to two similar-sized matrix, $x$ and $y$, can be defined mathematically as:

\begin{equation}
    SSIM(x,y) = \frac{(2\mu_x \mu_y + c_1)(2\sigma_{xy} + c_2)}{(\mu_x^2 + \mu_y^2 + c_1)(\sigma_x^2 + \sigma_y^2 + c_2)}
\end{equation}

where $\mu_x$ is the average of $x_i$, $\mu_y$ is the average of $y_i$, $c_1$ and $c_2$ are some constants, $\sigma_{xy}$ is the covariance of x and y, $\sigma_x^2$ and $\sigma_y^2$ are the variances of $x$ and $y$.

\subsection{Singular Value Decomposition}
In addition to the SSIM, another measure to compare matrices is using Singular Value Decomposition (SVD). A primary application of this in the current scenario is to compare the singular values, in decreasing order, to measure the so called 'energies' of the matrix. Mathematically for a matrix $\mathbf{M}$, this can be defined as:

\begin{equation}
    \mathbf{M = U \Sigma V^*}
\end{equation}

where $\mathbf{U}$ is a $m \times m$ complex unitary matrix, $\mathbf{\Sigma}$ is a $m \times n$ rectangular diagonal matrix with non-negative numbers on the diagonal (also known as singular values), and $\mathbf{V}$ is a $n \times n$ complex unitary matrix. The singular values $\mathbf{\Sigma1 \geq \Sigma2 \geq ... \geq \Sigma m \geq 0}$ are in descending order along the main diagonal of $\mathbf{\Sigma}$. The most important information or the highest 'energetic modes' are stored in the first few columns.

In the world of image processing, singular values have been used to compress the information in fewer bits using the first few singular values. In comparing the singular values for layer 4 connections in the different networks, the close similarities in the original network and the 50\% pruned network is encouraging. This essential means that the 50\% network retains most of the relevant information from the original network as evidenced in Figure \ref{fig:Histogram}. This in conjunction with a high degree of conservation of the saliency score (Table \ref{tab:ScoreSum}) shows that the 50\% pruning is most appropriate for network compression.

\end{document}